\documentclass[12pt]{iopart}

\usepackage{iopams}  
\usepackage{graphicx}
\begin{document}

\title[How to get from static to dynamic electromagnetism]{How to get from static to dynamic electromagnetism}

\author{J\"urgen K\"onig}

\address{Theoretische Physik and CENIDE, Universit\"at Duisburg-Essen, 47057 Duisburg, Germany}
\ead{koenig@thp.uni-due.de}
\vspace{10pt}
\begin{indented}
\item[]

\end{indented}

\begin{abstract}
We demonstrate how to derive Maxwell's equations, including Faraday's law and Maxwell's correction to Amp\`ere's law, by generalizing the description of static electromagnetism to dynamical situations.
Thereby, Faraday's law is introduced as a consequence of the relativity principle rather than an experimental fact, in contrast to the historical course and common textbook presentations. 
As a by-product, this procedure yields explicit expressions for the infinitesimal Lorentz and, upon integration, the finite Lorentz transformation.
The proposed approach helps to elucidate the relation between Galilei and Lorentz transformations and provides an alternative derivation of the Lorentz transformation without explicitly referring to the speed of light.
\end{abstract}

%
%
%
%
%

The theory of electromagnetism, usually taught after a course on mechanics, introduces as a new concept the electromagnetic field.
It is characterized by the electric and magnetic field strengths, denoted by $\mathbf{E}(\mathbf{r},t)$ and $\mathbf{B}(\mathbf{r},t)$.
Maxwell's famous equations describe how $\mathbf{E}(\mathbf{r},t)$ and $\mathbf{B}(\mathbf{r},t)$ arise from static and moving charges, expressed in terms of the charge density $\rho(\mathbf{r},t)$ and the charge current density $\mathbf{j}(\mathbf{r},t)$, but also how $\mathbf{E}(\mathbf{r},t)$ and $\mathbf{B}(\mathbf{r},t)$ influence each other.
The connection to mechanics is provided by the electromagnetic force on charged bodies, referred to as the Lorentz force.

Some textbooks on electrodynamics write down Maxwell's equations at the very beginning and, then, specify them when discussing special limits such as electrostatics and magnetostatics.
Others rather follow the historical course by starting with electro- and magnetostatics, then include Faraday's law as an experimental fact, and, finally, add Maxwell's correction to Amp\`ere's law to ensure charge conservation.
In both cases, the relativity principle is only discussed after treating electromagnetic waves in vacuum and demonstrating that their propagation speed $c = 1/\sqrt{\mu_0 \epsilon_0}$ is independent of the chosen inertial frame of reference.
The Lorentz transformation as the correct transformation between two inertial systems is, then, derived from combining the relativity principle with the universality of the  vacuum speed of light.
It couples time and space coordinates in some intricate way that leads to astonishing effects such as time dilatation and Lorentz contraction, which are at odds with our intuition built on everyday life.
To unify mechanics and electrodynamics within one consistent theory, one has to replace Newtonian mechanics by relativistic mechanics, and one ends up with Einstein's special relativity theory.

In this paper, we propose an alternative way to teach electrodynamics.
As sketched in Table~\ref{tab:three_steps}, Maxwell's equations are derived by extending the field equations of electro- and magnetostatics in three steps.
In the first step, all quantities are simply made time dependent.
The result is, however, in conflict with both charge conservation and the relativity principle.
To guarantee charge conservation, we add, as the second step, Maxwell's correction to Amp\`ere's law.
We show that the resulting field equations are Galilei invariant, while the Lorentz force is not.
Therefore, we require, in the third and final step, the relativity principle to hold for both the field equations and the Lorentz force.
This leads us to Faraday's law, which completes Maxwell's equations.
In addition, this procedure provides a recipe for how to correctly transform from one inertial system to another, first for an infinitesimal relative velocity of inertial systems and then, after integration, for a finite relative velocity.
The result is the Lorentz transformation.

\begin{table}
\caption{\label{tab:three_steps}
Static electromagnetism is described by the static field equations~(1) and the Lorentz force~(2). 
Dynamic electromagnetism is achieved in three steps.
First, all quantities are made time dependent, leading to the field equations~(3).
Second, charge conservation is used to obtain Maxwell's correction to Amp\`ere's law. 
This leads to the field equations~(6).
Finally, Faraday's law is obtained from the relativity principle.
The result is Maxwell's equations~(22).
}
\bigskip
	\begin{tabular*}{15.5cm}{cp{5.3cm}c}
	\hline\hline\\
	static electromagnetism & & dynamic electromagnetism \\
	 \hspace*{-.7cm} \vline & & \hspace*{1.3cm} \vline \\
	\end{tabular*}
	
	\begin{tabular*}{15.5cm}{p{1.2cm}p{1.5cm}p{1.5cm}p{1.5cm}p{1.5cm}p{1.5cm}p{1.5cm}p{1.5cm}p{1.5cm}}
	& (1) & $\Longrightarrow$ & (3) & $\Longrightarrow$ & (6) & $\Longrightarrow$ & (22)& \\
	\end{tabular*}
	
	\begin{tabular*}{15.5cm}{ccccc}
	\hspace*{1.9cm} & \vline & \hspace*{.15cm} \vline & \vline & \\
	& time dependence &  charge conservation &  \hspace*{.15cm} relativity principle\\
	\\
	\hline \hline
	\end{tabular*}
\end{table}

\section{From Static to Dynamic Electromagnetism in Three Steps}
\label{sec:from_static_to_dynamic}

The starting point for deriving Maxwell's equation is the description of static electromagnetism, summarized by the static field equations (in SI units)
\numparts
\begin{eqnarray}
	\label{M0a}
	\nabla \cdot \mathbf{E}(\mathbf{r}) &= \frac{\rho(\mathbf{r})}{\epsilon_0} 
	\\
	\label{M0b}
	\nabla \times \mathbf{E}(\mathbf{r}) &= \mathbf{0} 
	\\
	\label{M0c}
	\nabla \cdot \mathbf{B}(\mathbf{r}) &= 0
	\\
	\label{M0d}
	\nabla \times \mathbf{B}(\mathbf{r}) &= \mu_0 \mathbf{j}(\mathbf{r})
\end{eqnarray}
\endnumparts
as well as the Lorentz force
\begin{equation}
\label{L0}
	\mathbf{F} = Q (\mathbf{E} + \mathbf{v} \times \mathbf{B})
\end{equation}
that acts on a test charge $Q$ that moves with velocity $\mathbf{v}$.

In static electromagnetism, an electromagnetic field can be considered as the coexistence of two independent physical entities, namely an electric field generated by a static charge density $\rho(\mathbf{r})$ and a magnetic field generated by a stationary and divergence-free charge current density $\mathbf{j}(\mathbf{r})$ with $\nabla \cdot \mathbf{j}(\mathbf{r})=0$.
The respective source terms $\rho(\mathbf{r})$ and $\mathbf{j}(\mathbf{r})$ can be understood as being independent of each other (although physical currents are, of course, composed of moving charges).
On the one hand, pure electrostatics appears for a static charge distribution, such that $\mathbf{j}=\mathbf{0}$.
Pure magnetostatics, on the other hand, can be realized by a stationary current distribution composed of moving charges of one type (e.g., negatively charged electrons) while static charges of another type (e.g., positively charged ions) guarantee charge neutrality, such that $\rho=0$.
In that respect, \textit{static electromagnetism} is nothing more than a collective term for the two independent theories of electrostatics, described by Eqs.~(\ref{M0a}) and (\ref{M0b}), and magnetostatics, described by Eqs.~(\ref{M0c}) and (\ref{M0d}).
The Lorentz force~(\ref{L0}) on a test charge is, then, nothing but the superposition of the electric and magnetic force generated by the electric and magnetic field, respectively.

Since static electromagnetism is a special case of dynamic electromagnetism, it is impossible to deduce Maxwell's equations from the static field equations~(\ref{M0a})-(\ref{M0d}).
Therefore, we use inductive reasoning to derive Maxwell's equations by enlarging ad-hoc the applicability range of the field equations and, then, introducing those modifications enforced by the compatibility with fundamental principles, namely charge conservation and the relativity principle.

\subsection{First step: make $\rho$, $\mathbf{j}$, $\mathbf{E}$, and $\mathbf{B}$ time dependent}

In a first step, we simply make the charge density $\rho$, the current density $\mathbf{j}$, the electric field strength $\mathbf{E}$, and the magnetic field strength $\mathbf{B}$ time dependent. 
This leads to the field equations
\numparts
\begin{eqnarray}
	\label{M1a}
	\nabla \cdot \mathbf{E}(\mathbf{r},t) &=& \frac{\rho(\mathbf{r},t)}{\epsilon_0}
	\\
	\label{M1b}
	\nabla \times \mathbf{E}(\mathbf{r},t) &=& \mathbf{0}
	\\
	\label{M1c}
	\nabla \cdot \mathbf{B}(\mathbf{r},t) &=& 0
	\\
	\label{M1d}
	\nabla \times \mathbf{B}(\mathbf{r},t) &=& \mu_0 \mathbf{j}(\mathbf{r},t) \, .
\end{eqnarray}
\endnumparts
The Lorentz force is still given by Eq.~(\ref{L0}), but now with time-dependent electric and magnetic field strengths.

While the expression for the Lorentz force is already the correct one, the fields equation are not yet complete.
We nowadays know that the time derivative of both the electric and the magnetic field strength have to enter.
The incompleteness of the above field equations may be demonstrated experimentally. 
A much stronger argument to include them comes, however, from checking the consistency with fundamental principles. 

\subsection{Charge conservation}

One of these fundamental principles is charge conservation, expressed through the continuity equation
\begin{equation}
\label{continuity}
	\frac{\partial \rho (\mathbf{r},t)}{\partial t} + \nabla \cdot \mathbf{j}(\mathbf{r},t) =0 \, .
\end{equation}
To demonstrate that the field equations violate charge conservation, we take the divergence of Eq.~(\ref{M1d}) and use that $\mathrm{div} \, \mathrm{curl} \equiv 0$, so that 
\begin{equation}
	0 = \nabla \cdot ( \nabla \times \mathbf{B}) = \nabla \cdot ( \mu_0 \mathbf{j} ) = - \mu_0 \frac{\partial \rho}{\partial t} \neq 0  \, . 
\end{equation}

\subsection{Second step: add Maxwell's correction to Amp\`ere's law}

As noticed by Maxwell, the inconsistency with charge conservation can be cured by adding to the current density the so-called displacement current density,
$\mathbf{j} \mapsto \mathbf{j} + \epsilon_0 \frac{\partial \mathbf{E}}{\partial t}$.
This leads to the field equations
\numparts
\begin{eqnarray}
	\label{M2a}
	\nabla \cdot \mathbf{E} &=& \frac{\rho}{\epsilon_0}
	\\
	\label{M2b}
	\nabla \times \mathbf{E} &=& \mathbf{0}
	\\
	\label{M2c}
	\nabla \cdot \mathbf{B} &=& 0
	\\
	\label{M2d}
	\nabla \times \mathbf{B} &=& \mu_0 \mathbf{j} + \mu_0 \epsilon_0 \frac{\partial \mathbf{E}}{\partial t} 
\end{eqnarray}
\endnumparts
where the last term in Eq.~(\ref{M2d}) is called Maxwell's correction to Amp\`ere's law.
It fixes charge conservation since now
\begin{equation}
	0 = \nabla \cdot ( \nabla \times \mathbf{B}) = \nabla \cdot \left( \mu_0 \mathbf{j} + \mu_0 \epsilon_0 \frac{\partial \mathbf{E}}{\partial t} \right) = 
	\mu_0 \left( \nabla \cdot \mathbf{j} + \frac{\partial \rho}{\partial t} \right) = 0  \,\,\, \checkmark \, .
\end{equation}

In the static case, the field equations for the electric and the magnetic field strengths were independent of each other.
They could be split into two equations for electrostatics and two for magnetostatics, with $\rho(\mathbf{r})$ and $\mathbf{j}(\mathbf{r})$ being considered as independent quantities.
The generalization to the dynamic case makes all four field equations coupled to each other, since now the source terms $\rho(\mathbf{r},t)$ for the electric and $\mathbf{j}(\mathbf{r},t)$ for the magnetic field strength are connected to each other via the continuity equation.
In addition, Maxwell's correction to Amp\`ere's law introduces an explicit coupling between $\mathbf{E}(\mathbf{r},t)$ and $\mathbf{B}(\mathbf{r},t)$ for time-dependent electric field strengths.
What does it mean for the interpretation of the electromagnetic field?
At the present stage, we could still consider the electric and the magnetic field as two separate physical entities (a notion that we have to revise after the discussion of the relativity principle below), but now they are at least coupled to each other. 

In Eq.~(\ref{M2d}), the combination $\mu_0 \epsilon_0$ appears.
It is convenient to introduce the abbreviation
\begin{equation}
	\mu_0 \epsilon_0 = \frac{1}{c^2}
\end{equation}
where the constant $c$ has the dimension of a velocity.
This constant is identical to the vacuum speed of light derived from the full electrodynamic theory (which includes Faraday's law).
At the present stage, however, the field equations~(\ref{M2a})-(\ref{M2d}) do not support propagating solutions for the electromagnetic field in vacuum, i.e., the concept of light is so far meaningless. 
Nevertheless, $c$ provides a scale for expressing velocities such as the relative velocity between two inertial systems.

To arrive at Maxwell's equations, we need now to include Faraday's law.
Historically, the motivation for this came from experiment.
Here, we suggest, as an alternative, to derive Faraday's law from the relativity principle.

\subsection{Relativity principle}

The relativity principle states
\begin{center}
	\fbox{The laws of nature have the same form in all inertial systems.}
\end{center}
The relativity principle does \textit{not} make a statement of \textit{how} to transform from one inertial system to another.
If a candidate for such a transformation is given, one can \textit{check} whether the relativity principle is respected for some theory with the given transformation.
But one can also turn the argument the other way around and \textit{construct} the transformation by postulating the relativity principle to hold for the considered theory.
We will see both ways of reasoning in the following.

Within Newtonian mechanics, which is usually taught prior to electrodynamics, the situation is clear.
It respects the relativity principle when using the Galilei transformation to change between different inertial systems.
Newtonian mechanics is, therefore, called Galilei invariant.

\subsubsection{Galilei transformation.}

In the following, we denote by $S'$ an inertial system that moves with constant velocity $\mathbf{u}$ relative to the reference system $S$.
In both systems, any event is characterized by the respective time and space coordinates.
The Galilei transformation describes how the coordinates in $S'$ are connected to those in $S$.
It reads 
\numparts
\begin{eqnarray}
	t' &= t \\
	\mathbf{r}' &= \mathbf{r} - \mathbf{u} t \, .
\end{eqnarray}
\endnumparts
By writing $t (\mathbf{r}',t') = t'$ and $\mathbf{r} (\mathbf{r}',t') = \mathbf{r}' + \mathbf{u} t'$ as functions of $\mathbf{r}'$ and $t'$ and using the chain rule, we find $\frac{\partial}{\partial t'}=\frac{\partial t}{\partial t'}\frac{\partial}{\partial t} + \frac{\partial \mathbf{r}}{\partial t'}\cdot \nabla = \frac{\partial}{\partial t} + \mathbf{u} \cdot \nabla$ as well as $\frac{\partial}{\partial x'}=\frac{\partial t}{\partial x'}\frac{\partial}{\partial t} + \frac{\partial \mathbf{r}}{\partial x'}\cdot \nabla = \frac{\partial}{\partial x}$ and similarly for $y$ and $z$.
This implies the transformation rules
\numparts
\begin{eqnarray}
	\frac{\partial}{\partial t'} & = \frac{\partial}{\partial t} + \mathbf{u} \cdot \nabla \\
	\nabla' &= \nabla
\end{eqnarray}
\endnumparts
for the derivatives with respect to time and space.

\subsubsection{Newtonian mechanics is Galilei invariant.}

For Newtonian mechanics to be Galilei invariant, the equation of motion $\mathbf{F} = m \ddot{\mathbf{r}}$ has to keep its form under a Galilei transformation.
The Galilei transformation implies $\ddot{\mathbf{r}}'= \ddot{\mathbf{r}}$ for the acceleration of a body.
If we, further, assume that the mass of the body is a universal property for all inertial systems, $m'=m$, then Galilei invariance of the equation of motion implies 
\begin{equation}
\label{F'=F}
	\mathbf{F}' = \mathbf{F} \, ,
\end{equation}
i.e., the force on a body has the same \textit{value} in all inertial systems.
We note that Eq.~(\ref{F'=F}) is a stronger statement than mere form invariance of the equations of motions.
Form invariance of an equation is, per definition, guaranteed by the relativity principle, irrespective of which transformation between inertial systems is considered. 
In contrast, the equal values of the force in all inertial systems is a special property of the Galilei transformation.

\subsubsection{Is electromagnetism Galilei invariant?}

The relativity principle holds for mechanics, and in favor of a unified description of nature we demand that it also holds for electromagnetism.
Since Newtonian mechanics is Galilei invariant, we check whether or not electromagnetism can be formulated in a Galilei-invariant way.
This includes not only the field equations~(\ref{M2a})-(\ref{M2d}) but also the Lorentz force~(\ref{L0}), which connects electromagnetism to mechanics.

The source terms for the electric and magnetic field strengths are the charge density and the current density.
Under Galilei transformation they transform according to
\numparts
\begin{eqnarray}
	\rho'(\mathbf{r}',t') &= \rho(\mathbf{r},t) \\
	\mathbf{j}'(\mathbf{r}',t') &= \mathbf{j}(\mathbf{r},t) - \mathbf{u} \rho(\mathbf{r},t) \, .
\end{eqnarray}
\endnumparts
As a consequence, the continuity equation is Galilei invariant, which is proven by
\begin{equation}
	\frac{\partial \rho'}{\partial t'} + \nabla' \cdot \mathbf{j}' = \left( \frac{\partial}{\partial t} + \mathbf{u} \cdot \nabla \right) \rho + \nabla \cdot \left( \mathbf{j} - \mathbf{u} \rho \right)= \frac{\partial \rho}{\partial t} + \nabla \cdot \mathbf{j} = 0 \,\,\, \checkmark \, .
\end{equation}

We now turn to the field equations~(\ref{M2a})-(\ref{M2d}).
By postulating their Galilei invariance, we can deduce how the electric and magnetic field strengths transform.
Since $\nabla' = \nabla$ and $\rho'=\rho$, it is obvious that Eqs.~(\ref{M2a}) and (\ref{M2b}) are Galilei invariant if the electric field strength has the same value in all inertial systems, $\mathbf{E}'=\mathbf{E}$.
To find the transformation behavior of the magnetic field strength, we rewrite Amp\`ere's law~(\ref{M2d}) as
\begin{eqnarray}
	\nabla \times \mathbf{B} &=& \mu_0 \mathbf{j} + \mu_0 \epsilon_0 \frac{\partial \mathbf{E}}{\partial t}
	\nonumber \\
	&=& \mu_0 \left( \mathbf{j}' + \mathbf{u} \rho \right) + \mu_0 \epsilon_0 \left( \frac{\partial }{\partial t'} - \mathbf{u} \cdot \nabla \right) \mathbf{E}
	\nonumber \\
	&=& \mu_0 \mathbf{j}' + \mu_0 \epsilon_0 \frac{\partial \mathbf{E}'}{\partial t'} + \mu_0 \epsilon_0 
	\underbrace{\left[ \mathbf{u} (\nabla \cdot \mathbf{E}) - (\mathbf{u} \cdot \nabla) \mathbf{E} \right]}_{\nabla \times (\mathbf{u} \times \mathbf{E} )} \, .
\end{eqnarray}
From 
\begin{equation}
	\nabla' \times \underbrace{\left( \mathbf{B} - \frac{\mathbf{u}}{c^2} \times \mathbf{E} \right) }_{\mathbf{B}'  } = \mu_0 \mathbf{j}'  + \mu_0 \epsilon_0 \frac{\partial \mathbf{E}'}{\partial t'}
\end{equation}
we read off the expression for $\mathbf{B}'$. 
Finally, we check that also the remaining field equation~(\ref{M2c}) is Galilei invariant,
\begin{eqnarray}
	\nabla' \cdot \mathbf{B}' &=& \nabla \cdot \left( \mathbf{B} - \frac{\mathbf{u}}{c^2} \times \mathbf{E} \right)
	= \underbrace{\nabla \cdot \mathbf{B}}_{0} - \frac{1}{c^2} \nabla \cdot (\mathbf{u} \times \mathbf{E}) 
	\nonumber \\
	&=& \frac{1}{c^2} \mathbf{u} \cdot  \underbrace{(\nabla  \times \mathbf{E})}_{\mathbf{0}} = 0 \qquad \checkmark \, .
\end{eqnarray}

To summarize, we find that the field equations Eqs.~(\ref{M2a})-(\ref{M2d}) are, indeed, Galilei invariant if the field strengths transform according to
\numparts
\begin{eqnarray}
	\label{E_Galilei}
	\mathbf{E}'(\mathbf{r}',t') &= \mathbf{E}(\mathbf{r},t) \\
	\label{B_Galilei}
	\mathbf{B}'(\mathbf{r}',t')&= \mathbf{B}(\mathbf{r},t) - \frac{\mathbf{u}}{c^2} \times \mathbf{E}(\mathbf{r},t) \, . 
\end{eqnarray}
\endnumparts

An immediate consequence of Eq.~(\ref{B_Galilei}) is that the value of the magnetic field strength depends on the chosen inertial system.
It is, therefore, no longer possible to consider the electric and the magnetic field as two separate physical entities.
The electric and the magnetic field strength $\mathbf{E}$ and $\mathbf{B}$ should rather be viewed as two facets of the very same physical entity, namely the \textit{electromagnetic field}.

We have seen that not only Newtonian mechanics but also the field equations~(\ref{M2a})-(\ref{M2d}) of the electromagnetic field (which do not include Faraday's law) are Galilei invariant.
Does it mean we have found a consistent theory that unifies mechanics and electromagnetism in a Galilei-invariant way?
No, it doesn't.
What remains to be checked is the \textit{connection} between mechanics and electromagnetism, provided by the Lorentz force.
And here comes the trouble: the Lorentz force is \textit{not} Galilei invariant.

To transform the Lorentz force~(\ref{L0}) on a test charge $Q$ that moves with velocity $\mathbf{v}$ in $S$ into another inertial system $S'$ that moves with $\mathbf{u}$ relative to $S$, we make use of $Q'=Q$, $\mathbf{v}'=\mathbf{v}-\mathbf{u}$, $\mathbf{F}'=\mathbf{F}$, $\mathbf{E}'=\mathbf{E}$, and $\mathbf{B}'= \mathbf{B} - \frac{\mathbf{u}}{c^2} \times \mathbf{E}$ to find
\begin{equation}
	\mathbf{F}' = Q' \left( \mathbf{E}' + \mathbf{v}' \times \mathbf{B}' \right)
	+ Q' \left[ \mathbf{u} \times \mathbf{B}' + (\mathbf{v}'+\mathbf{u}) \times \left( \frac{\mathbf{u}}{c^2} \times \mathbf{E}' \right) \right]
\end{equation}
which clearly differs in its form from Eq.~(\ref{L0}).
The terms containing $\mathbf{u}$ obviously break Galilei invariance.

Another way to drastically demonstrate the breakdown of Galilei invariance is to consider a point charge $Q$ that is moving with velocity $\mathbf{v}$ in a static magnetic field $\mathbf{B}$ that is generated by a magnet at rest.
In the rest frame $S$ of the magnet, there is a finite Lorentz force $\mathbf{F} = Q \mathbf{v} \times \mathbf{B}$.
If we switch to the (momentary) rest frame $S'$ of the point charge with the help of the Galilei transformation with $\mathbf{u} = \mathbf{v}$, we find that neither the electric field (due to $\mathbf{E}'=\mathbf{E} = \mathbf{0}$) nor the magnetic field (due to $\mathbf{v}'=\mathbf{0}$) contributes to the Lorentz force, hence $\mathbf{F}'=\mathbf{0}$, in contradiction to Eq.~(\ref{F'=F}).
Not only are the values for the force different in different inertial systems.
The force is even vanishing in one system while being finite in another one.
So, depending on the chosen inertial system we come to different conclusions of whether the point charge remains in uniform motion or not.
This clearly contradicts the relativity principle.

There are two possibilities of how to deal with this dilemma.
The pessimistic option is to give up the relativity principle, i.e., to accept that there is one distinguished inertial system and the correct force on a point charge is the Lorentz force calculated in this special inertial system.
To have a special inertial system in electromagnetism is in line with the notion of an ether, which was a popular concept in the nineteenth century.
The electromagnetic field is, then, considered as an excitation of the ether, and there is no reason to expect the field equations to be valid in any inertial system different from the rest frame of the ether.

The optimistic option is to take this dilemma as a challenge to modify the theory such that the relativity principle remains valid.
It might be tempting to simply make the expression for the Lorentz force Galilei invariant.
This, however, would mean to completely drop the magnetic force contribution $Q \mathbf{v}\times\mathbf{B}$.
The advantage of achieving a consistent, Galilei-invariant, unified description of both mechanics and electromagnetism would, thus, come with the drawback of totally removing magnetism: the magnetic field strength $\mathbf{B}$ would still be present in the field equations but it would have no effect whatsoever, neither on charged bodies nor on the electric field strength.
A world without magnetic phenomena, however, is \textit{not} the world we are living in.

If the Lorentz force is not changed, then the relativity principle can only be saved by modifying the transformation that connects different inertial systems, i.e., the Galilei transformation has to be replaced by some other transformation, which we call, by definition, the Lorentz transformation.
The task is now to derive the explicit expressions of the Lorentz transformation by imposing the requirement that electromagnetism is Lorentz invariant.
As we will see below, following this route yields Faraday's law as a necessary consequence of the relativity principle.
At the same time, it leads in a natural and transparent way to explicit expressions for the Lorentz transformation.

\subsubsection{Infinitesimal Lorentz transformation of the electromagnetic field.}

The Lorentz transformation involves a so-far unknown $\mathbf{u}$-dependence of all the quantities $t'$, $\mathbf{r}'$, $\frac{\partial}{\partial t'}$, $\nabla'$, $\rho'$, $\mathbf{j}'$, $\mathbf{E}'$, and $\mathbf{B}'$.
For an infinitesimally small $\mathbf{u}$, an expansion up to linear order is sufficient.
We refer to this linear expansion as the infinitesimal Lorentz transformation.
While for half of these quantities, namely $\mathbf{r}'$, $\frac{\partial}{\partial t'}$, $\mathbf{j}'$, and $\mathbf{B}'$, the correct linear term is already provided by the Galilei transformation, we need to determine the corresponding linear terms for the other half, namely $t'$, $\nabla'$, $\rho'$, and $\mathbf{E}'$.

We start to derive the infinitesimal Lorentz transformation of the electric and magnetic field strength.
For this, we consider an inertial system $S$ which hosts a static electromagnetic field with field strengths $\mathbf{E}$ and $\mathbf{B}$ and a point charge $Q$ that moves with infinitesimal velocity $\mathbf{u}$.
The force on the point charge is $\mathbf{F} = Q (\mathbf{E} + \mathbf{u} \times \mathbf{B})$.
In the (momentary) rest frame $S'$ of the point charge, the force is $\mathbf{F}' = Q \mathbf{E}'$.
By requiring 
\begin{equation}
	\mathbf{F}' = \mathbf{F} + \mathcal{O} \left( \frac{u^2}{c^2} \right)
\end{equation}
and understanding Eq.~(\ref{B_Galilei}) as the linear approximation of the finite Lorentz transformation of the magnetic field strength, we obtain for the infinitesimal Lorentz transformation 
\numparts
\begin{eqnarray}
	\mathbf{E}' &= \mathbf{E} + \mathbf{u} \times \mathbf{B} + \mathcal{O}\left( \frac{u^2}{c^2} \right)\\
	\mathbf{B}' &= \mathbf{B} - \frac{\mathbf{u}}{c^2} \times \mathbf{E} + \mathcal{O}\left( \frac{u^2}{c^2} \right) \, .
\end{eqnarray}
\endnumparts
By inductive reasoning, we assume that this transformation behavior is not restricted to the case of static $\mathbf{E}$ and $\mathbf{B}$, but valid in general.

We remark that, in contrast to the Galilei transformation, not only the magnetic but also the electric field strength depends on the chosen inertial system.
This confirms once more that the electromagnetic field should not be considered as a mere coexistence of two separate physical entities.

Since we have changed, as compared to the Galilei transformation, the transformation behavior of the electromagnetic field, we cannot expect the field equations~(\ref{M2a})-(\ref{M2d}) to be form invariant under the new transformation.
In fact, invariance under the new transformation enforces a modification of Eqs.~(\ref{M2b}) that leads to Faraday's law.

\subsubsection{Faraday's law.}

Let us consider a static magnetic field, i.e., $\frac{\partial \mathbf{B}}{\partial t} = \mathbf{0}$, in the absence of an electric field, $\mathbf{E}=\mathbf{0}$.
Applying the infinitesimal Lorentz transformation yields $\mathbf{E}' = \mathbf{u} \times \mathbf{B} + \mathcal{O} (\frac{u^2}{c^2})$ and $\mathbf{B}' = \mathbf{B} + \mathcal{O}(\frac{u^2}{c^2})$.
Using $\nabla' = \nabla + \mathcal{O}(\frac{u}{c})$ as well as $\frac{\partial}{\partial t'} = \frac{\partial}{\partial t} + \mathbf{u} \cdot \nabla + \mathcal{O}( \frac{u^2}{c^2} )$, we find
\begin{eqnarray}
	\nabla' \times \mathbf{E}' &= \nabla' \times \left( \mathbf{u} \times \mathbf{B} \right) +  \mathcal{O}\left( \frac{u^2}{c^2} \right) 
	= \nabla \times \left( \mathbf{u} \times \mathbf{B} \right) +  \mathcal{O}\left( \frac{u^2}{c^2} \right) \nonumber \\
	&=\mathbf{u}\underbrace{\left( \nabla \cdot \mathbf{B} \right)}_{0} - \underbrace{\left( \mathbf{u} \cdot \nabla \right)}_{\frac{\partial}{\partial t'} - \frac{\partial}{\partial t}+ \mathcal{O}\left(\frac{u^2}{c^2}\right)} \mathbf{B} +  \mathcal{O}\left( \frac{u^2}{c^2} \right) \nonumber \\
	&= \underbrace{\frac{\partial \mathbf{B}}{\partial t}}_{\mathbf{0}} - \frac{\partial \mathbf{B}}{\partial t'} +  \mathcal{O}\left( \frac{u^2}{c^2} \right)
	\nonumber \\
	&= - \frac{\partial \mathbf{B}'}{\partial t'} +  \mathcal{O}\left( \frac{u^2}{c^2} \right) \, .
\end{eqnarray}
This is nothing but Faraday's law, derived for an inertial system that moves infinitesimally slow with respect to a system in which only a static magnetic field exists.
Again, we use inductive reasoning to assume that the relation $\nabla \times \mathbf{E} = - \frac{\partial \mathbf{B}}{\partial t}$ is generally valid also beyond this specific scenario.

In conclusion, Faraday's law appears as a by-product in trying to save the relativity principle for electromagnetism. 
There is no need to rely on experimental input at this stage.

\subsection{Third step: add Faraday's law}

By including Faraday's law, we arrive at the final form of the field equations, the celebrated Maxwell equations
\numparts
\begin{eqnarray}
	\label{M3a}
	\nabla \cdot \mathbf{E} &=& \frac{\rho}{\epsilon_0}
	\\
	\label{M3b}
	\nabla \times \mathbf{E} &=& - \frac{\partial \mathbf{B}}{\partial t} 
	\\
	\label{M3c}
	\nabla \cdot \mathbf{B} &=& 0
	\\
	\label{M3d}
	\nabla \times \mathbf{B} &=& \mu_0 \mathbf{j} + \mu_0 \epsilon_0 \frac{\partial \mathbf{E}}{\partial t}  \, .
\end{eqnarray}
\endnumparts 

Faraday's law makes the field equations more symmetric with respect to the electric and the magnetic field strength.
Not only does a time-dependent electric field strength $\mathbf{E}$ serve as a source of a magnetic field strength $\mathbf{B}$, but also a time-dependent magnetic field strength $\mathbf{B}$ generates an electric field strength $\mathbf{E}$.
This mutual interdependence of $\mathbf{E}$ and $\mathbf{B}$ is an import prerequisite for propagating solutions of the electromagnetic field.
In that respect, we have arrived at a truly \textit{dynamic} theory of electromagnetism, usually referred to as \textit{electrodynamcis}.

What still needs to be checked, of course, is that all four of Maxwell's equations are Lorentz invariant.
We will do that in the next section by applying an infinitesimal Lorentz transformation.
Thereby, we make use of the freedom to choose the so-far undetermined linear terms (in $\mathbf{u}/c$) of the transformed charge density $\rho'$, the gradient $\nabla'$, and the time $t'$ such that the relativity principle is fulfilled for electrodynamics. 
Since time and space define the stage not only for electrodynamics but also for mechanics, we finally end up with a consistent theory for both of them.

\section{Infinitesimal Lorentz transformation}
\label{sec:infinite_Lorentz}

\subsection{Infinitesimal Lorentz transformation of time and space coordinates}

First, we postulate the Lorentz invariance of $\nabla \cdot \mathbf{B} = 0$ to derive how the gradient transforms.
Combining
\begin{eqnarray}
	0 &= \nabla \cdot \mathbf{B} = \nabla \cdot \left( \mathbf{B}' + \frac{\mathbf{u}}{c^2} \times \mathbf{E} \right)
	+  \mathcal{O}\left( \frac{u^2}{c^2} \right)
	\nonumber \\
	&= \nabla \cdot \mathbf{B}' - \frac{\mathbf{u}}{c^2} \cdot \underbrace{\left( \nabla \times \mathbf{E} \right)}_{-\frac{\partial \mathbf{B}}{\partial t}}
	+  \mathcal{O}\left( \frac{u^2}{c^2} \right)
	= \underbrace{\left( \nabla + \frac{\mathbf{u}}{c^2} \frac{\partial }{\partial t} \right)}_{\nabla' + \mathcal{O}\left( \frac{u^2}{c^2} \right) } \cdot \mathbf{B}' +  \mathcal{O}\left( \frac{u^2}{c^2} \right)
\end{eqnarray}
with the postulate $\nabla' \cdot \mathbf{B}' = 0$ fixes the expression for $\nabla'$.
Therefore, the derivatives with respect to space and time transform according to 
\numparts
\begin{eqnarray}
	\frac{\partial}{\partial t'} &= \frac{\partial}{\partial t} + \mathbf{u} \cdot \nabla +  \mathcal{O}\left( \frac{u^2}{c^2} \right)\\
	\nabla' &= \nabla + \frac{\mathbf{u}}{c^2} \frac{\partial}{\partial t} +  \mathcal{O}\left( \frac{u^2}{c^2} \right) \, . 
\end{eqnarray}
\endnumparts 
 
Next, we derive the transformation of the coordinates themselves.
From $\frac{\partial t}{\partial t'} = (\frac{\partial}{\partial t} + \mathbf{u} \cdot \nabla) t +  \mathcal{O}(\frac{u^2}{c^2})= 1+  \mathcal{O}(\frac{u^2}{c^2})$ as well as $\nabla' t = (\nabla + \frac{\mathbf{u}}{c^2} \frac{\partial}{\partial t}) t +  \mathcal{O}(\frac{u^2}{c^2})= \frac{\mathbf{u}}{c^2}+  \mathcal{O}(\frac{u^2}{c^2})$, we get $t=t' +\frac{\mathbf{u}\cdot \mathbf{r}'}{c^2}+  \mathcal{O}(\frac{u^2}{c^2})$ and, therefore,
\numparts
\begin{eqnarray}
	t' &=& t - \frac{\mathbf{u} \cdot \mathbf{r}}{c^2} + \mathcal{O}\left( \frac{u^2}{c^2} \right)\\
	\mathbf{r}' &=& \mathbf{r} - \mathbf{u} t + \mathcal{O}\left( \frac{u^2}{c^2} \right) \, .
\end{eqnarray}
\endnumparts 
In contrast to the Galilei transformation, time remains no longer invariant under a change of the inertial system.

\subsection{Infinitesimal Lorentz transformation of charge and current density}

Next, we postulate Lorentz invariance of Gau{\ss}'s law.
This leads to
\begin{eqnarray}
	\frac{\rho}{\epsilon_0} &=& \nabla \cdot \mathbf{E} 
	= \left( \nabla' - \frac{\mathbf{u}}{c^2} \frac{\partial}{\partial t} \right) \cdot 
	\left( \mathbf{E}' - \mathbf{u} \times \mathbf{B} \right)
	+  \mathcal{O}\left( \frac{u^2}{c^2} \right)
	\nonumber \\
	&=& \nabla' \cdot \mathbf{E}' - \frac{\mathbf{u}}{c^2} \cdot \frac{\partial \mathbf{E}'}{\partial t}
	-\nabla' \cdot \left( \mathbf{u} \times \mathbf{B} \right)
	+  \mathcal{O}\left( \frac{u^2}{c^2} \right)
	\nonumber \\
	&=& \frac{\rho'}{\epsilon_0} - \frac{\mathbf{u}}{c^2} \cdot \frac{\partial \mathbf{E}}{\partial t}
	- \underbrace{\nabla \cdot \left( \mathbf{u} \times \mathbf{B} \right)}_{-\mathbf{u} \cdot (\nabla \times \mathbf{B})}
	+  \mathcal{O}\left( \frac{u^2}{c^2} \right)
	\nonumber \\
	&=& \frac{\rho'}{\epsilon_0} 
	+ \mathbf{u}\cdot \underbrace{\left( \nabla \times \mathbf{B} -\frac{1}{c^2}\frac{\partial \mathbf{E}}{\partial t} \right)}_{\mu_0 \mathbf{j}}
	+  \mathcal{O}\left( \frac{u^2}{c^2} \right)
\end{eqnarray}
from which we deduce
\numparts
\begin{eqnarray}
	\rho' &=& \rho - \frac{\mathbf{u}\cdot \mathbf{j}}{c^2} +  \mathcal{O}\left( \frac{u^2}{c^2} \right)\\
	\mathbf{j}' &=& \mathbf{j} - \mathbf{u} \rho +  \mathcal{O}\left( \frac{u^2}{c^2} \right) \, .
\end{eqnarray}
\endnumparts 
As a consequence, the value of the charge density depends on the chosen inertial system.
It is not constant, as the Galilei transformation would suggest.

\subsection{Lorentz invariance of Faraday's and Amp\`ere's law}

Finally, we check the Lorentz invariance of Faraday's and Amp\`ere's law.
We get
\begin{eqnarray}
	\nabla' \times \mathbf{E}' + \frac{\partial \mathbf{B}'}{\partial t'}
	&=& 
	\left( \nabla + \frac{\mathbf{u}}{c^2} \frac{\partial}{\partial t} \right) \times
	\left( \mathbf{E} + \mathbf{u} \times \mathbf{B} \right)
	\nonumber \\
	&&
	+ \left( \frac{\partial}{\partial t} + \mathbf{u} \cdot \nabla \right)
	\left(  \mathbf{B} - \frac{\mathbf{u}}{c^2} \times \mathbf{E} \right) + \mathcal{O}\left( \frac{u^2}{c^2} \right)
	\nonumber \\
	&=& \underbrace{\nabla \times \mathbf{E} + \frac{\partial \mathbf{B}}{\partial t} }_{\mathbf{0}}
	+ \underbrace{ \left[ \nabla \times \left( \mathbf{u} \times \mathbf{B} \right) + \left( \mathbf{u}\cdot \nabla \right) \mathbf{B} \right]}_{\mathbf{u} \left( \nabla \cdot \mathbf{B} \right) = \mathbf{0}}
	+ \mathcal{O}\left( \frac{u^2}{c^2} \right)
	\nonumber \\
	&=& \mathcal{O}\left( \frac{u^2}{c^2} \right) \qquad \checkmark
\end{eqnarray}
as well as 
\begin{eqnarray}
	\nabla' \times \mathbf{B}' - \mu_0 \mathbf{j}' - \frac{1}{c^2} \frac{\partial \mathbf{E}'}{\partial t'}
	&=& 
	\left( \nabla + \frac{\mathbf{u}}{c^2} \frac{\partial}{\partial t} \right) \times
	\left( \mathbf{B} - \frac{\mathbf{u}}{c^2} \times \mathbf{E} \right) - \mu_0 \mathbf{j} + \mu_0 \mathbf{u} \rho
	\nonumber \\
	&&
	- \frac{1}{c^2} \left( \frac{\partial}{\partial t} + \mathbf{u} \cdot \nabla \right)
	\left(  \mathbf{E} + \mathbf{u} \times \mathbf{B}  \right)
	+ \mathcal{O}\left( \frac{u^2}{c^2} \right)
	\nonumber \\
	&=& \underbrace{\nabla \times \mathbf{B} - \mu_0 \mathbf{j} - \frac{1}{c^2} \frac{\partial \mathbf{E}}{\partial t}}_{\mathbf{0}}
	\nonumber \\
	&&
	+ \mu_0 \mathbf{u} \rho
	- \frac{1}{c^2}\underbrace{\left[ \nabla \times \left( \mathbf{u} \times \mathbf{E} \right) + \left( \mathbf{u}\cdot \nabla \right) \mathbf{E} \right]}_{\mathbf{u} \left( \nabla \cdot \mathbf{E} \right)=\mathbf{u} \rho/\epsilon_0}
	+ \mathcal{O}\left( \frac{u^2}{c^2} \right) 
	\nonumber \\
	&=& \mathcal{O}\left( \frac{u^2}{c^2} \right) \qquad \checkmark \, .
\end{eqnarray}

In summary, we have shown that Maxwell's equations as well as the Lorentz force are Lorentz invariant.
Furthermore, we have derived the infinitesimal Lorentz transformation of $t'$, $\mathbf{r}'$, $\frac{\partial}{\partial t'}$, $\nabla'$, $\rho'$, $\mathbf{j}'$, $\mathbf{E}'$, and $\mathbf{B}'$.
The results are summarized in Table~\ref{tab:Galilei_Lorentz}.

\begin{table}[h]
\caption{\label{tab:Galilei_Lorentz}
Comparison of the Galilei transformation with the infinitesimal Lorentz transformation.
Here, $\mathbf{u}$ denotes the velocity of inertial system $S'$ with respect to inertial system $S$.
}
\bigskip

	\begin{tabular}{lp{2.cm}l}
	\hline \hline
	Galilei transformation & & infinitesimal Lorentz transformation \\
	\hline \noalign{\vskip 2mm} 
	$t'=t$ & & $\displaystyle t' = t - \frac{\mathbf{u} \cdot \mathbf{r}}{c^2} + \mathcal{O}\left( \frac{u^2}{c^2} \right)$\\
	$\mathbf{r}' = \mathbf{r} - \mathbf{u} t $  & & $\displaystyle \mathbf{r}' = \mathbf{r} - \mathbf{u} t + \mathcal{O}\left( \frac{u^2}{c^2} \right)$ \\
	\noalign{\vskip 2mm}  \hline \noalign{\vskip 2mm} 
	$\displaystyle \frac{\partial}{\partial t'} = \frac{\partial}{\partial t} + \mathbf{u} \cdot \nabla$ & &  $\displaystyle \frac{\partial}{\partial t'} = \frac{\partial}{\partial t} + \mathbf{u} \cdot \nabla +  \mathcal{O}\left( \frac{u^2}{c^2} \right)$\\
	$\nabla' = \nabla $ & & $\displaystyle \nabla' = \nabla + \frac{\mathbf{u}}{c^2} \frac{\partial}{\partial t} +  \mathcal{O}\left( \frac{u^2}{c^2} \right)$\\
	\noalign{\vskip 2mm}  \hline \noalign{\vskip 2mm} 
	$\rho' = \rho$ & & $\displaystyle \rho' = \rho - \frac{\mathbf{u}\cdot \mathbf{j}}{c^2} +  \mathcal{O}\left( \frac{u^2}{c^2} \right)$\\
	$\mathbf{j}' = \mathbf{j} - \mathbf{u} \rho $ & & $\displaystyle \mathbf{j}' = \mathbf{j} - \mathbf{u} \rho +  \mathcal{O}\left( \frac{u^2}{c^2} \right)$\\
	\noalign{\vskip 2mm}  \hline \noalign{\vskip 2mm} 
	$\mathbf{E}' = \mathbf{E}$ & & $\displaystyle \mathbf{E}' = \mathbf{E} + \mathbf{u} \times \mathbf{B} + \mathcal{O}\left( \frac{u^2}{c^2} \right)$\\
	$\displaystyle \mathbf{B}' = \mathbf{B} - \frac{\mathbf{u}}{c^2} \times \mathbf{E}$ & & 
	$\displaystyle \mathbf{B}' = \mathbf{B} - \frac{\mathbf{u}}{c^2} \times \mathbf{E} + \mathcal{O}\left( \frac{u^2}{c^2} \right)$\\
	\noalign{\vskip 2mm} 
	\hline\hline
	\end{tabular}
\end{table}

\section{Finite Lorentz transformation}
\label{sec:finite_Lorentz}

Based on the results for the infinitesimal Lorentz transformation, we derive the finite Lorentz transformation by repeatedly applying infinitesimal ones.
In order to keep the notation transparent, we choose the boost direction (i.e., the direction of the relative velocity between two inertial systems) from now on always along the $x$-axis.
Since the $y$- and $z$-components (i.e., those \textit{perpendicular} to the boost direction) of $\mathbf{r}$, $\nabla$, and $\mathbf{j}$ remain unchanged under an infinitesimal Lorentz transformation, they also remain unchanged under a finite one.
The same holds true to the $x$-components (i.e., the one \textit{parallel} to the boost direction) of the electric and magnetic field strength $\mathbf{E}$ and $\mathbf{B}$.
The other components ($x$, $\frac{\partial}{\partial x}$, $j_x$ as well as $E_y$, $E_z$, $B_y$, $B_z$), on the other hand, do change under Lorentz transformation, and we need to determine how.

We start with the Lorentz transformation of time and space coordinates and their derivatives.
Afterwards, we determine the transformation of the charge and current density and, finally, of the electric and the magnetic field strength.

\subsection{Lorentz transformation of time and space coordinates and their derivatives}

Let $S$, $S'$, and $S''$ be three inertial systems, where $S'$ moves with finite velocity $u \, \hat{\mathbf{e}}_x$ relative to $S$, $S''$ moves with infinitesimal velocity $du \, \hat{\mathbf{e}}_x$ relative to $S'$, and the relative velocity of $S''$ with respect to $S$ is denoted by $\tilde u \, \hat{\mathbf{e}}_x$.
The finite Lorentz transformation from $S$ to $S'$ can be written in a vector-matrix notation
\begin{equation}
	\left( \begin{array}{c} ct' \\ x'  \end{array}\right)
	=
	\Lambda(u) \left( \begin{array}{c} ct \\ x \end{array} \right)
\end{equation}
by introducing the Lorentz-transformation matrix $\Lambda(u)$.
Since $ct$ and $x$ have the same units, the matrix elements of $\Lambda(u)$ are all dimensionless.
Similarly, $\Lambda(\tilde u)$ connects the coordinates specified in inertial systems $S''$ and $S$, respectively, and $\Lambda(du) + \mathcal{O}(\frac{(du)^2}{c^2})$ those in inertial systems $S''$ and $S'$.
The relation
\begin{equation}
\label{group}
	\Lambda(\tilde u) = \Lambda(du) \Lambda(u) + \mathcal{O}\left( \frac{(du)^2}{c^2} \right) \, .
\end{equation}
describes that the finite Lorentz transformation from $S$ to $S''$ can be decomposed into a finite one from $S$ to $S'$ and an infinitesimal one from $S'$ to $S''$.

From Section~\ref{sec:infinite_Lorentz}, we know how to perform the infinitesimal Lorentz transformation.
This is expressed by the symmetric matrix
\begin{equation}
	\Lambda(du) = \left( \begin{array}{cc} 1 & -\frac{du}{c} \\ -\frac{du}{c} & 1 \end{array} \right) + \mathcal{O}\left( \frac{(du)^2}{c^2} \right) 
\end{equation}
which we plug into Eq.~(\ref{group}) to get
\begin{equation}
\label{Lambda_tilde_1}
	\Lambda (\tilde u) = \Lambda(u) + \frac{du}{c} \left( \begin{array}{cc} 0 & -1 \\ -1 & 0 \end{array} \right) \Lambda(u)+ \mathcal{O}\left( \frac{(du)^2}{c^2} \right) \, .
\end{equation}

To proceed, we need to know how $\tilde u$ depends on $u$ and $du$. 
For this, we consider a particle that is at rest in $S$.
In $S'$, it has velocity $v'=\frac{dx'}{dt'}=-u$ and in $S''$, the velocity is given by $v'' = \frac{dx''}{dt''}=-\tilde u$.
This yields
\begin{eqnarray}
	\tilde u &=& - \frac{d x''}{dt''} = - \frac{d (x' - t' du)}{d (t'-\frac{x' du}{c^2})} + \mathcal{O}\left( \frac{(du)^2}{c^2} \right) =  \frac{- d x' + dt' du}{d t' - \frac{dx' du}{c^2}} + \mathcal{O}\left( \frac{(du)^2}{c^2} \right)
	\nonumber \\
	&=& \frac{u+du}{1+\frac{u du}{c^2}}+ \mathcal{O}\left( \frac{(du)^2}{c^2} \right) = (u+du) \left(1-\frac{u du}{c^2}\right) + \mathcal{O}\left( \frac{(du)^2}{c^2} \right)
	\nonumber \\
	&=& u + \left( 1-\frac{u^2}{c^2}\right) du + \mathcal{O}\left( \frac{(du)^2}{c^2} \right)  \, .
	\label{u_tilde}
\end{eqnarray}
We note that $\tilde u\neq u + du$ for finite $u$, i.e., the Galilean rule of adding velocities is no longer valid.
Furthermore, at this stage, the role of $c$ as a limiting velocity becomes apparent: for $u=c$, the extra Lorentz boost by $du$ does not increase the velocity anymore.
It is, thus, impossible to achieve $\tilde u > c$, i.e., velocities larger than the speed of light.

Plugging Eq.~(\ref{u_tilde}) into $\Lambda(\tilde u)$ and expanding up to linear order in $du$ yields
\begin{equation}
\label{Lambda_tilde_2}
	\Lambda (\tilde u) = \Lambda(u) + \left( 1-\frac{u^2}{c^2} \right) \frac{d\Lambda}{du} du + \mathcal{O}\left( \frac{(du)^2}{c^2} \right) \, .
\end{equation}
From combining Eqs.~(\ref{Lambda_tilde_1}) and (\ref{Lambda_tilde_2}), we obtain the differential equation
\begin{equation}
\label{DGL_u}
	\frac{d\Lambda}{d \left[\mathrm{arctanh} \left(\frac{u}{c}\right)\right]} = \left( \begin{array}{cc} 0 & -1 \\ -1 & 0 \end{array} \right) \Lambda \, , 
\end{equation}
for the Lorentz-transformation matrix $\Lambda$, where we have made use of the derivative $\frac{d}{dx} [\mathrm{arctanh} \, x] = \frac{1}{1-x^2}$.

It is quite intriguing that in Eq.~(\ref{DGL_u}) the velocity $u$ appears only in the combination 
\begin{equation}
	\theta = \mathrm{arctanh} \left(\frac{u}{c}\right) \, .
\end{equation}
This suggest a change of variable from the \textit{velocity} $u$ to the \textit{rapidity} $\theta$.
In terms of the rapidity, the differential equation for $\Lambda$ simplifies to
\begin{equation}
	\frac{d\Lambda}{d\theta}  = \left(\begin{array}{cc} 0 & -1 \\ -1 & 0  \end{array}\right) \Lambda \, .
\end{equation}
Instead of solving this first-order differential equation, it is convenient to calculate the second derivative.
We make use of $\left(\begin{array}{cc} 0 & -1 \\ -1 & 0  \end{array}\right) \left(\begin{array}{cc} 0 & -1 \\ -1 & 0  \end{array}\right) = \mathbf{1}$ and end up with the second-order differential equation
\begin{equation}
	\frac{d^2\Lambda}{d\theta^2}  = \Lambda 
\end{equation}
that is easy to solve.
Each matrix element of $\Lambda$ is a linear combination of 
\begin{equation}
	\cosh \theta = \frac{1}{\sqrt{1-\tanh^2\theta}} =  \frac{1}{\sqrt{1-\frac{u^2}{c^2}}} = \gamma
\end{equation}
and
\begin{equation}
	\sinh \theta = \frac{\tanh \theta}{\sqrt{1-\tanh^2\theta}} =  \frac{\frac{u}{c}}{\sqrt{1-\frac{u^2}{c^2}}} = \gamma \frac{u}{c}
\end{equation}
where $\gamma = 1 / \sqrt{1-\frac{u^2}{c^2}}$ denotes the famous Lorentz factor.

The coefficients of the linear combinations of $\cosh \theta$ and $\sinh \theta$ for the matrix elements of $\Lambda$ have to be chosen such that the expansion up to linear order in $\theta$ is given by the infinitesimal Lorentz transform.
Because of $\cosh \theta = 1 + \mathcal{O}(\frac{u^2}{c^2})$ and $\sinh \theta = \frac{u}{c} + \mathcal{O}(\frac{u^2}{c^2})$, this results in 
\begin{equation}
	\Lambda = \left(\begin{array}{cc} \cosh \theta & - \sinh \theta \\ -\sinh \theta & \cosh \theta 
	\end{array}\right)
		= \gamma \left(\begin{array}{cc} 1 & - \frac{u}{c} \\ - \frac{u}{c} & 1 \end{array}\right) \, .
\end{equation}
In comparison to the infinitesimal Lorentz transformation, the finite one is obtained by simply multiplying with the Lorentz factor $\gamma$ in the time coordinate and the spatial coordinate along the boost direction.
In conclusion, the final result for the finite Lorentz transformation is
\numparts
\begin{eqnarray}
	t' &=&\gamma \left( t - \frac{u x}{c^2} \right) \\
	x' &=& \gamma \left( x - ut \right)  \, .
\end{eqnarray}
\endnumparts

To determine the transformation of the derivatives with respect to time and spatial coordinates, we write $t(x',t')=\gamma \left( t'+\frac{ux'}{c^2} \right)$ and $x(x',t)=\gamma(x'+ut')$ as functions of $x'$ and $t'$ and use the chain rule for derivatives.
This leads to
\numparts
\begin{eqnarray}
	\frac{\partial}{\partial t'} &=&\gamma \left( \frac{\partial}{\partial t} + u \frac{\partial}{\partial x} \right) \\
	\frac{\partial}{\partial x'} &=& \gamma \left( \frac{\partial}{\partial x} + 
		\frac{u}{c^2} \frac{\partial}{\partial t} \right)  \, .
\end{eqnarray}
\endnumparts
So, again, the finite Lorentz transformation differs from the infinitesimal one just by the Lorentz factor $\gamma$ in the derivatives of the time coordinate and the spatial coordinate along the boost direction.

\subsection{Lorentz transformation of charge and current density}

The infinitesimal Lorentz transformation of $\rho$ and $\mathbf{j}$ is identical to the one of $t$ and $\mathbf{r}$.
Therefore, also the finite Lorentz transformations have to be identical.
This leads to the result
\numparts
\begin{eqnarray}
	\rho' &=&\gamma \left( \rho - \frac{u j_x}{c^2} \right) \\
	j'_x &=& \gamma \left( j_x - u \rho \right)
\end{eqnarray}
\endnumparts
which, again, differs from the infinitesimal transformation only by the Lorentz factor $\gamma$ in the charge density and the current density along the boost direction.

\subsection{Lorentz transformation of electric and magnetic field strength}

Finally, we consider the electric and magnetic field strength.
The field strengths along the boost direction remain unchanged.
For the perpendicular directions, we define the Lorentz-transformation matrix $M(u)$ by
\begin{equation}
	\left( \begin{array}{c} E_y'  \\ E_z' \\ c B_y' \\ c B_z' \end{array}\right)
	= M(u) \left( \begin{array}{c} E_y \\ E_z \\ c B_y \\ c B_z \end{array} \right)
\end{equation}
Since $E$ and $cB$ have the same units, the matrix elements of $M(u)$ are all dimensionless. 
Analogous to the procedure for the time and space coordinates, we derive a differential equation for $M(u)$ from
\begin{equation}
	M (\tilde u) = M(du) M(u) + \mathcal{O}\left( \frac{(du)^2}{c^2} \right) 
\end{equation}
by using the infinitesimal Lorentz transformation
\begin{equation}
	M(du) =
	\left( \begin{array}{cccc}  1 & 0 & 0  & -\frac{du}{c} \\ 0 & 1 & \frac{du}{c} & 0 \\ 0 & \frac{du}{c} & 1 &0 \\ -\frac{du}{c} & 0  & 0 & 1 \end{array} \right)
		+ \mathcal{O}\left( \frac{(du)^2}{c^2} \right) \, 
\end{equation}
then expressing $\tilde u$ in terms of $u$ and $du$ through Eq.~(\ref{u_tilde}), and finally performing the variable change from velocity to rapidity.
This results in the first-order differential equation
\begin{equation}
	\frac{d M}{d\theta}  =  \left( \begin{array}{cccc} 0 & 0 & 0 & -1 \\ 0 & 0 & 1 & 0 \\ 0 & 1 & 0 & 0 \\ -1 & 0 & 0 & 0 \end{array} \right) M \, .
\end{equation}
Again, it is convenient to calculate the second derivative.
This leads to the second-order differential equation 
\begin{equation}
	\frac{d^2 M}{d\theta^2} = M
\end{equation}
which is easily solved by linear combinations of $\cosh \theta$ and $\sinh \theta$ in each matrix element.
The coefficients of these linear combination are, again, fixed by comparison with the infinitesimal Lorentz transformation.
This yields the final result
\numparts
\begin{eqnarray}
	\mathbf{E}'_\perp &=& \gamma \left( \mathbf{E}_\perp + \mathbf{u} \times \mathbf{B}_\perp \right) \\
	\mathbf{B}'_\perp &=& \gamma \left( \mathbf{B}_\perp - \frac{\mathbf{u}}{c^2} \times \mathbf{E}_\perp \right)
\end{eqnarray}
\endnumparts
where the $y$- and $z$-components of the electric and magnetic field strength are combined into $\mathbf{E}_\perp$ and $\mathbf{B}_\perp$.
Once more, the finite Lorentz transformation differs from the infinitesimal one only by the Lorentz factor $\gamma$ in the perpendicular field-strength components.

The comparison between the infinitesimal and the finite Lorentz transformation for the boost direction along the $x$-axis is summarized in Table~\ref{tab:inf_finite_Lorentz}.

\begin{table}
\caption{\label{tab:inf_finite_Lorentz}
Comparison of infinitesimal with finite Lorentz transformation.
Here, $u$ denotes the velocity of inertial system $S'$ with respect to inertial system $S$ along the $x$-direction.
}
\bigskip

	\begin{tabular}{lp{2.cm}l}
	\hline \hline
	infinitesimal Lorentz transformation & & finite Lorentz transformation \\
	\hline \noalign{\vskip 2mm} 
	$\displaystyle t' = t - \frac{u x}{c^2} + \mathcal{O}\left( \frac{u^2}{c^2} \right)$ & & $\displaystyle t' =\gamma \left( t - \frac{u x}{c^2} \right)$\\
	$\displaystyle x' = x - ut + \mathcal{O}\left( \frac{u^2}{c^2} \right)$  & & $\displaystyle x' =\gamma \left( x - u t \right)$ \\
	$y' = y$ & & $y'=y$\\
	$z' = z$ & & $z'=z$\\
	\noalign{\vskip 2mm}  \hline \noalign{\vskip 2mm} 
	$\displaystyle \frac{\partial}{\partial t'} =\frac{\partial}{\partial t} + u \frac{\partial}{\partial x} + \mathcal{O}\left( \frac{u^2}{c^2} \right)$ & & $\displaystyle \frac{\partial}{\partial t'} =\gamma \left( \frac{\partial}{\partial t} + u \frac{\partial}{\partial x} \right)$
	\medskip
	\\
	$\displaystyle \frac{\partial}{\partial x'} =\frac{\partial}{\partial x} + \frac{u}{c^2} \frac{\partial}{\partial t} + \mathcal{O}\left( \frac{u^2}{c^2} \right)$ & & $\displaystyle \frac{\partial}{\partial x'} =\gamma \left( \frac{\partial}{\partial x} + \frac{u}{c^2} \frac{\partial}{\partial t} \right)$ 
	\medskip
	\\
	$\displaystyle \frac{\partial}{\partial y'} = \frac{\partial}{\partial y}$ & &  $\displaystyle \frac{\partial}{\partial y'} = \frac{\partial}{\partial y}$
	\medskip
	\\
	$\displaystyle \frac{\partial}{\partial z'} = \frac{\partial}{\partial z}$ & &  $\displaystyle \frac{\partial}{\partial z'} = \frac{\partial}{\partial z}$
	\\
	\noalign{\vskip 2mm}  \hline \noalign{\vskip 2mm} 
	$\displaystyle \rho' = \rho - \frac{u j_x}{c^2} + \mathcal{O}\left( \frac{u^2}{c^2} \right)$ & & $\displaystyle \rho' =\gamma \left( \rho - \frac{u j_x}{c^2} \right)$\\
	$\displaystyle j_x' = j_x - u\rho + \mathcal{O}\left( \frac{u^2}{c^2} \right) $  & & $\displaystyle j_x' =\gamma \left( j_x - u \rho \right)$ \\
	$j_y' = j_y$ & & $j_y'=j_y$\\
	$j_z' = j_z$ & & $j_z'=j_z$\\
	\noalign{\vskip 2mm}  \hline \noalign{\vskip 2mm} 
	$E_x' = E_x$ & & $E_x' = E_x$\\
	$\displaystyle \mathbf{E}'_\perp = \mathbf{E}_\perp + \mathbf{u} \times \mathbf{B}_\perp + \mathcal{O}\left( \frac{u^2}{c^2} \right)$ & &  
	$\displaystyle \mathbf{E}'_\perp = \gamma \left( \mathbf{E}_\perp + \mathbf{u} \times \mathbf{B}_\perp \right)$\\
	$B_x' = B_x$ & & $B_x' = B_x$\\
	$\displaystyle \mathbf{B}'_\perp = \mathbf{B}_\perp - \frac{\mathbf{u}}{c^2} \times \mathbf{E}_\perp + \mathcal{O}\left( \frac{u^2}{c^2} \right)$ & &  
	$\displaystyle \mathbf{B}'_\perp = \gamma \left( \mathbf{B}_\perp - \frac{\mathbf{u}}{c^2} \times \mathbf{E}_\perp \right)$\\
	\noalign{\vskip 2mm} 
	\hline\hline
	\end{tabular}
\end{table}

\section{Discussion}

\subsection{Relation between the Galilei and the Lorentz transformation}

In the presented approach, the Lorentz transformation has been constructed out of the Galilei transformation via the infinitesimal Lorentz transformation as an intermediate step.
To clarify the relation between the Galilei and the Lorentz transformation, we now go the opposite way (as it is usually done in the literature) and ask how to obtain the Galilei transformation as a limiting case of the Lorentz transformation.

In contrast to what is sometimes claimed in textbooks, the Galilei transformation is \textit{not} just the low-velocity limit of the Lorentz transformation.
It is true that in the low-velocity limit, $u\ll c$, the Lorentz factor $\gamma$ drops out because of $\gamma = 1 +\mathcal{O}(\frac{u^2}{c^2})$.
The linearization of the Lorentz transformation in $u/c$ does, however, not lead to the Galilei transformation but rather to the infinitesimal Lorentz transformation, see Table~\ref{tab:inf_finite_Lorentz}.
The Galilei transformation is only achieved after dropping the term linear in $u$ in the transformations of time, spatial derivative, charge density, and electric field strength, see Table~\ref{tab:Galilei_Lorentz}.
This means that, in addition to the low-velocity limit, another approximation is involved.
Neglecting the linear term in the transformation of the time coordinate means that the time difference of two events considered within this approximation must be taken much larger than the spatial difference, $c\Delta t \gg |\Delta \mathbf{r}|$, a condition referred to as \textit{largely timelike} \cite{Levy-Leblond,Bellac}.
To make this approximation consistent, also the linear term in the transformation of the electric field strength has to be neglected, i.e., the magnetic field strength must be much smaller than the electric one.
For this reason, the description of the electromagnetic field within this approximation, given by Eqs.~(\ref{M2a})-(\ref{M2d}), has been called the \textit{electric limit} \cite{Bellac,Montigny,Heras,Rousseaux}.
As discussed in Section~\ref{sec:from_static_to_dynamic}, the problem with this Galilean description of electromagnetism is that, in order to be consistent, also the term linear in $u$ in the Lorentz force, i.e., the magnetic force needs to be neglected.
As a consequence, the magnetic field strength does not have any impact.

Dropping the linear terms in the transformations of time, spatial derivative, charge density, and electric field strength but keeping them in the transformations of spatial coordinate, time derivative, current density, and magnetic field strength seems quite arbitrary.
It is interesting to note that one can also do it just the other way around.
In this case, the time difference of two events considered within this approximation must be much smaller than the spatial difference, $c\Delta t \ll |\Delta \mathbf{r}|$, a condition referred to as \textit{largely spacelike} \cite{Levy-Leblond,Bellac}.
Furthermore, the magnetic field strength must be much larger than the electric one, which motivates the term \textit{magnetic limit} \cite{Bellac,Montigny,Heras,Rousseaux} for the resulting description of electromagnetism.
In the magnetic limit, there is no problem with the Lorentz force but now the continuity equation is not satisfied, as discussed in Ref.~\cite{Bellac}.

\subsection{What makes electrodynamics relativistic?}

Science aims at describing \textit{how} nature is and not \textit{why} it is the way it is.
It is, therefore, useless to ask why electrodynamics is a relativistic theory.
Nevertheless, it is interesting from the pedagogical point of view trying to identify the origin of relativistic behavior of electrodynamics, which may be considered as an answer to the \textit{why} question.

Before doing so, however, we need to clarify what is meant by \textit{relativistic}.
For historical reasons, electrodynamics is usually called \textit{relativistic}, while Newtonian mechanics is referred to as \textit{nonrelativistic}.
This nomenclature is, however, very unfortunate because, as a matter of fact, both theories are in accord with the relativity principle.
The distinction between the two theories consists in \textit{how} to transform from one inertial system to another. 
We, therefore, prefer to call electrodynamics \textit{Lorentz relativistic} and Newtonian mechanics \textit{Galilei relativistic}.
So, the question we pose here is what makes electrodynamics Lorentz relativistic instead of Galilei relativistic.

The crucial feature of the Lorentz transformation is the existence of a \textit{fixed} velocity scale, i.e., a special velocity $c$ that is determined by physical constants or processes.
Lorentz transformations with different values of $u/c$ have very different appearances.
The Galilei transformation, on the other hand, is scale free, i.e., there is no special velocity.
Of course, if one wants to compare temporal and spacial coordinates with each other then some velocity scale is needed also for the Galilei transformation.
But this velocity scale is not fixed.
It can, rather, be chosen arbitrarily, without changing the transformation itself.
In this sense, we identify the origin of Lorentzian relativity of a theory through the existence of a \textit{fixed} velocity scale.

In most textbook presentations, the Lorentz transformation is introduced after discussing propagating electromagnetic waves in vacuum, i.e., light.
According to Maxwell's equations, the vacuum speed of light is universal and given by the constant $c=1/\sqrt{\mu_0\epsilon_0}$.
The Lorentz-relativistic nature of electrodynamics is, thus, based on the properties of a dynamic phenomenon, namely the propagation of light.
While there is nothing wrong with this interpretation, the approach presented in this paper suggests an alternative way to explain why electrodynamics must be Lorentz relativistic, but now based on \textit{static} electromagnetism.

In our derivation of Maxwell's equation, Galilei invariance had to be abandoned during the third step only because the Lorentz force is not Galilei invariant.
The Lorentz force on a charged body consists of an electric and a magnetic part. 
The electric force is independent of and the magnetic force is proportional to the velocity of the body. 
In order to compare electric with magnetic forces or to add them to one resulting force, there must be some fixed velocity scale.
This conclusion is, of course, independent of the chosen unit system.

A nice illustration of the velocity scale $c=1/\sqrt{\mu_0\epsilon_0}$ in \textit{static} electromagnetism is the situation studied in Problem 5.13 of Ref.~\cite{Griffiths}.
There, two positively charged, infinitely long, infinitely thin, parallel wires are considered, which move together with constant velocity $v$ along the wires' direction.
In this scenario, both the charge density and the current density are time independent, leading to time-independent electric and magnetic field strengths.
Because of the charges, there is an electrostatic repulsive force between the wires.
In addition, since the moving charges constitute a current, there is an attractive magnetostatic force. 
It is a simple exercise of electro- and magnetostatics to calculate the ratio of the magnetic over the electric force.
The ratio turns out to be $v^2/c^2$.
As a consequence, it vanishes for $v=0$ and approaches unity for $v=c$.
Therefore, $c$ is the velocity limit at which the magnetic force between the two wires compensates the electric one.

In conclusion, the very existence of a magnetic force in addition to an electric one can, therefore, be viewed as the origin of the Lorentz-relativistic nature of electrodynamics.

\subsection{How to teach electrodynamics}

Most textbook presentations of electrodynamics follow the historical course:
after treating electro- and magnetostatics, Faraday's law is introduced as an experimental fact and, then, the magnetostatic version of Amp\`ere's law is modified by adding Maxwell's correction.
We suggest to change this modus operandi in two respects.

First, the order of introducing Faraday's law and Maxwell's correction should be interchanged. 
The advantage of doing so has already been emphasized in Ref.~\cite{Jammer}.
Second, we suggest to derive Faraday's law from the relativity principle rather than introducing it as an experimental fact.
This has, to the best of our knowledge, not yet been discussed in the literature so far. 

There are several advantages of the proposed approach.
To begin with, Faraday's law is not introduced as an experimental fact but rather as a necessity to comply with fundamental principles.
This gives students a stronger justification for why Faraday's law looks like the way it does.
Furthermore, the proposed order of steps allows us to discuss the principle of relativity at an earlier stage than it is usually done.
The pedagogical virtue of this discussion is that it better connects Newtonian mechanics to electrodynamics, it highlights the principle of relativity as a fundamental concept, and it employs the Galilei transformation which is more intuitive and better accessible to students than the Lorentz transformation.
The practical exercise of deriving the transformation behavior of the electric and magnetic field strength under a Galilei transformation provides an opportunity to become acquainted with the relativity principle before dealing with the complications associated with the Lorentz transformation.
Furthermore, the important insight that the electric and magnetic field are not two separate physical entities but two facets of the very same physical entity, namely the electromagnetic field, is easier to digest when being illustrated with the help of the Galilei rather than the Lorentz transformation.

For an introductory course on electrodynamics, we suggest to include the material presented in Section~\ref{sec:from_static_to_dynamic} but to skip Sections~\ref{sec:infinite_Lorentz} and \ref{sec:finite_Lorentz} since they might be too cumbersome for novice students.
The author has made this decision for the last two rounds of teaching electrodynamics to third semester physics students and found it appropriate and well-balanced.
The derivation of the infinitesimal and the finite Lorentz transformation in Sections~\ref{sec:infinite_Lorentz} and \ref{sec:finite_Lorentz} is rather meant for more advanced students and for instructors.
Its pedagogical value lies in the fact that the infinitesimal Lorentz transformation obtained in Section~\ref{sec:infinite_Lorentz} has a less complicated structure than the finite one, making it easier to digest.
This advantage has also been emphasized in Ref.~\cite{Galili} (where the infinitesimal Lorentz transformation is referred to as the weak relativistic approximation).
The commonalities and differences with the Galilei transformation are also better visible for the infinitesimal Lorentz transformation than for the finite one.
Also the integration from the infinitesimal to the finite Lorentz transformation in Section~\ref{sec:finite_Lorentz} is of pedagogical value, as it naturally introduces the notion of rapidity.
Finally, deriving the Lorentz transformation without explicitly referring to the speed of light helps to identify the origin of the (Lorentz-)relativistic nature of electrodynamics, namely the very existence of a magnetic force. 

\section{Acknowledgements.}

We thank Eric Kleinherbers for drawing our attention to Ref.~\cite{Jammer} and for helpful discussions.

\end{document}